\documentclass{aa}
\usepackage{graphicx}
\usepackage{amsmath}
\usepackage{natbib}
\usepackage{hyperref}

\titlerunning{}

\usepackage{color}

\usepackage{cleveref}
\newcommand{\lcar}{\mbox{l}~Car}

\newcommand{\kms}{\mbox{$\mbox{km\,s}^{-1}$}}

\begin{document}

\title{CRIRES high-resolution infrared spectroscopy of the long-period Cepheid {\mbox{l}~Car}\thanks{\lcar(lowercase~l)$\equiv$HD48410 should not  be confused  with L~Car(uppercase~L)$\equiv$HD~90264 or  i~Car(lowercase~i)$\equiv$HD~79447 or I~Car(uppercase~I)$\equiv$HD~90589 or $\iota$~Car(iota)$\equiv$HD~80404. To avoid these misidentifications,  $\ell$~Car  \citep[e.g.,][]{nardetto09,anderson16} or $l$~Car \citep[e.g.,][]{neilson16} have been used instead in the recent  literature.}}
\titlerunning{CRIRES high-resolution infrared spectroscopy of {\mbox{l}~Car}}
\authorrunning{Nardetto et al. }

\author{N.~Nardetto \inst{1} 
\and E.~Poretti \inst{2,3} 
\and A.~Gallenne\inst{4}  
\and M.~Rainer \inst{2} 
\and R.I.~Anderson \inst{5}
\and P. Fouqu\'e\inst{6}  
\and W.~Gieren\inst{7, 8}  
\and D.~Graczyk \inst{7, 8, 9} 
\and P.~Kervella \inst{10,11}  
\and P.~Mathias\inst{12, 13}
\and A.~M\'erand \inst{5} 
\and D.~Mourard \inst{1}  
\and H.~Neilson\inst{14} 
\and G.~Pietrzynski \inst{9} 
\and B.~Pilecki \inst{9}  
\and J.~Storm \inst{15}  
\and S.~Borgniet \inst{10}  
\and A. Chiavassa  \inst{1} 
\and V. Hocd\'e  \inst{1} 
\and B. Trahin \inst{10}  
}
\institute{Universit\'e C\^ote d'Azur, OCA, CNRS, Lagrange, France,  Nicolas.Nardetto@oca.eu   
\and  INAF -- Osservatorio Astronomico di Brera, Via E. Bianchi 46, 23807 Merate (LC), Italy 
\and  Fundaci\'{o}n Galileo Galilei -- INAF, Rambla Jos\'{e} Ana Fernandez P\'{e}rez 7, 38712 -- Bre\~{n}a
Baja, Spain
\and European Southern Observatory, Alonso de C\'ordova 3107, Casilla 19001, Santiago, Chile 
\and European Southern Observatory, Karl-Schwarzschild-Str. 2, D-85748 Garching b. M\"unchen, Germany
\and Observatoire Midi-Pyr\'en\'ees, Laboratoire d'Astrophysique, UMR 5572, Universit\'e Paul Sabatier - Toulouse 3, 14 avenue Edouard Belin, 31400 Toulouse, France  
\and Departamento de Astronom\'ia, Universidad de Concepci\'on, Casilla 160-C, Concepci\'on, Chile 
\and Millenium Institute of Astrophysics, Santiago, Chile 
\and Nicolaus Copernicus Astronomical Center, Polish Academy of Sciences, ul. Bartycka 18, PL-00-716 Warszawa, Poland
\and LESIA (UMR 8109), Observatoire de Paris, PSL, CNRS, UPMC, Univ. Paris-Diderot, 5 place Jules Janssen, 92195 Meudon, France 
\and Unidad Mixta Internacional Franco-Chilena de Astronom\'ia, CNRS/INSU, France (UMI 3386) and Departamento de Astronom\'ia, Universidad de Chile, Camino El Observatorio 1515, Las Condes, Santiago, Chile 
 \and   Universit\'e de Toulouse, UPS-OMP, Institut de Recherche en Astrophysique et Plan\'etologie, Toulouse, France
\and CNRS, UMR5277,  Institut de recherche en Astrophysique et Plan\'etologie, 14 Avenue Edouard Belin, 31400 Toulouse, France 
\and Department of Astronomy \& Astrophysics, University of Toronto, 50 St. George Street, Toronto, ON, M5S 3H4 
\and Leibniz Institute for Astrophysics, An der Sternwarte 16, 14482, Potsdam, Germany 
}

\date{Received ... ; accepted ...}

\abstract{The dynamical structure of the atmosphere of Cepheids has been well studied in the optical. 
Several authors have found very interesting spectral features in the J band, but little data have been secured beyond 1.6~$\mu$m.  
However, such observations 
can probe different radial velocities and line asymmetry regimes, 
and are able to provide crucial insights into stellar physics.}
{Our  goal was to investigate the infrared line-forming region in the K-band domain, and its impact on the projection factor and the $k$-term of Cepheids.} 
{ We secured CRIRES observations for the long-period Cepheid \lcar, with a focus on the unblended spectral line \ion{NaI} 2208.969~nm. We measured  the corresponding radial velocities (by using the first moment method) and the line asymmetries (by using the bi-Gaussian method). These quantities are compared to the HARPS visible spectra  we previously obtained on \lcar.} 
{The optical and infrared radial velocity curves show the same amplitude (only about 3\% of difference), with a slight radial velocity shift of about $0.5\pm0.3$~\kms\ between the two curves. Around the minimum radius (phase $\simeq$ 0.9) the visible radial velocity curve is found in advance compared to the infrared one (phase lag), which is consistent with an infrared line forming higher in the atmosphere (compared to the visible line) and with a compression wave moving from the bottom to the top of the atmosphere during maximum outward velocity. The asymmetry of the K-band line is also found to be significantly different from that of the optical line.
} 
{}


\maketitle

\section{Introduction}\label{s_Introduction}



For almost a century, the Baade--Wesselink (BW) method is used to derive the distance of Cepheids  
\citep{lindermann18,baade26, wesselink46}.  The concept is simple: distances are computed using measurements of the angular diameter over the whole pulsation period along with the stellar radius variations deduced from the 
integration of the pulsation velocity $V_{\rm puls}$. The pulsation velocity is linked to the observed radial velocity (RV)
by the projection factor $p=V_{\rm puls}/RV$  \citep{nardetto04, nardetto17}. There are three versions of 
the BW method corresponding to different ways of determining the angular diameter curve: a photometric 
version based on infrared surface brightness relations \citep{fouque97, fouque07, storm11a, storm11b}; an 
interferometric version \citep{lane00, kervella04a, merand05}; and a more recent version that combines 
several photometric bands, velocimetry, and interferometry \citep[SPIPS; ][]{merand15}. 
On the other hand, the radial velocity curve is usually 
derived from optical spectroscopy. This means that 
the projection factor, which is wavelength-dependent \citep{nardetto09, neilson12}, is also calculated in the visible. Infrared spectroscopy can also be used to derive the radius variation, but the projection factor must be calculated consistently, 
i.e.,  taking into account both the expected limb darkening of the star in the infrared domain, and  
the atmospheric velocity gradient as probed by the infrared spectral lines. 

On the interferometric side, Cepheids are usually observed at infrared wavelengths, whereas at optical wavelengths the pulsation has actually been resolved for only three stars, $\beta$~Dor, $\eta$~Aql (A.P. Jacob 2008 PhD \footnote{\url{http://www.physics.usyd.edu.au/~ande/Thesis_PhD_APJ.pdf}}), and the long-period \lcar\ \citep{davis09}. For the latter, the angular diameter curves derived respectively from the infrared surface brightness relation and infrared interferometry are consistent \citep{kervella04d}, while it is not the case for the short-period Cepheid $\delta$~Cep \citep{ngeow12}. Interestingly, a resolved structure around $\delta$ Cep has  recently been discovered in the visible spectral range using interferometry \citep{nardetto16a}. 

\begin{figure*}[htbp]
\begin{center}
\resizebox{0.7\hsize}{!}{\includegraphics[clip=true]{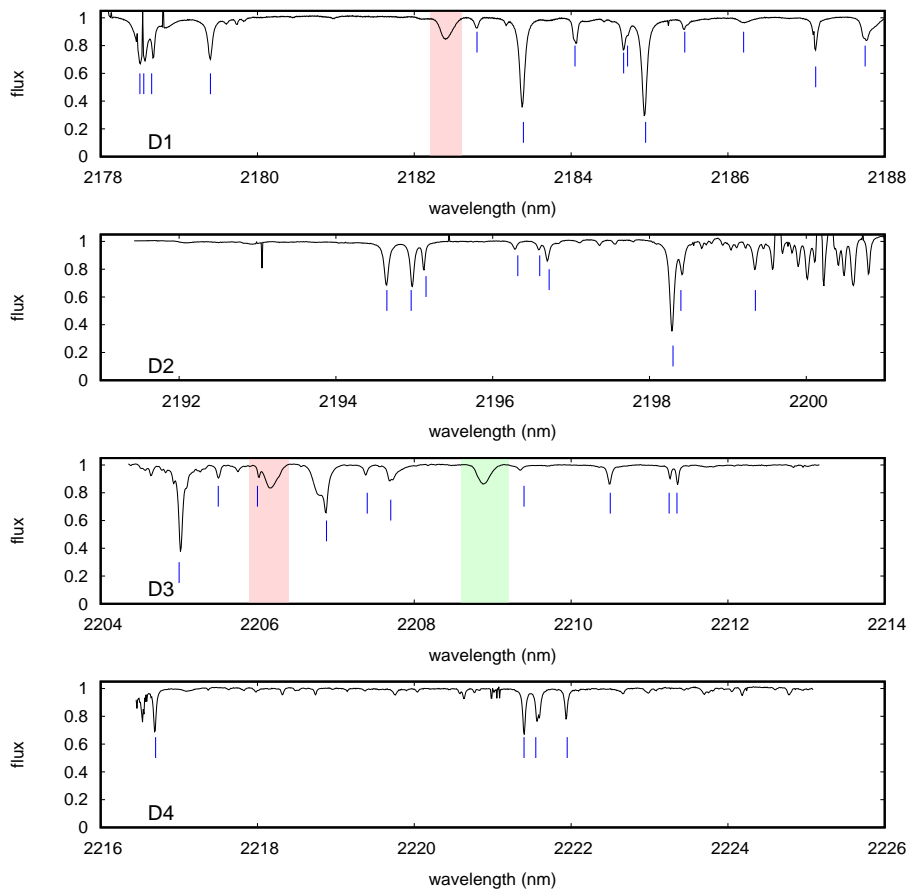}}
\end{center}
\caption{CRIRES spectra corresponding to phase 0.0 are plotted over the four detectors D1 to D4. The blue vertical features indicate the telluric lines as identified with the Molecfit ESO sky modeling tool. The atmospheric line indicated on D1 by a red region is blended between phase 0.6 to 1.0. The \ion{NaI}  doublet appears in D3:  The first component (2206.242~nm, red region) is blended, while the second (2208.969~nm, green region) is analyzed in this paper.} \label{Fig.spectra}
\end{figure*}

In these three BW approaches an underlying assumption is that the size of the star (i.e., its photosphere, independently of its limb darkening) has the same position whatever the wavelength domain considered. 
\citet{sasselov90} used the static models from \citet{kurucz79} and derived the depth of the photosphere (defined by an optical depth of $\frac{2}{3}$ in the continuum) as a function of the wavelength (their Fig. 21). The impact of this effect on the derived distance was estimated to be around 1\%\ or 2\%\  by \citet{nardetto11b} using a hydrodynamical model of \lcar. Another interesting piece of evidence comes from \citet{anderson16} who found different motions of \lcar\, by tracing its pulsation with contemporaneous optical velocimetry and infrared interferometry.

The dynamical structure of the atmosphere of Cepheids has been well studied in the optical 
\citep{sanford56, bell64, karp75c, sasselov90,  wallerstein15, nardetto06a, anderson14, anderson16, nardetto17}, while only a few spectroscopic observations are available in the infrared. At 1.1~$\mu$m, \citet{sasselov89} 
used the Fourier Transform Spectrometer (FTS) at the Canada-France-Hawaii Telescope (CFHT) 
\citep{maillard82} to monitor two Cepheids, X~Sgr ($P=$7.013~d) and $\eta$~Aql ($P=$7.177~d). They reported RV 
curves with systematically larger amplitudes (20--35\%) in the infrared than in the optical domain. 
This result was later confirmed in a subsequent study based on a larger set of infrared 
spectroscopic data \citep{sasselov90}. In addition, \citet{sasselov94a, sasselov94b, sasselov94c} 
studied the HeI 10830~\AA\,  spectral line to model the chromospherical structure of  
Cepheids.

This paper  presents the very first analysis of  infrared spectra
in the K band (2.2~$\mu$m) of the Cepheid \lcar. The final goal is  to investigate the behavior of the RV curve
at such long wavelengths.


\begin{table*}
\begin{center}
\caption[]{Log and results of the CRIRES observations of \lcar. 
Phases and cycles are computed with T$_0$=JD~2456023.533 and $P=35.5496$ d.
Cycles are counted from the first observed one. The signal-to-noise ratio (S/N) is calculated over the wavelength range 2212.0--2212.3 nm for 2012 and 2013 data, while the range 2208.2--2208.5 nm is used for the 2008 data. 
\label{Tab_res}}
\setlength{\doublerulesep}{\arrayrulewidth}
\begin{tabular}{ccccrrccrc}
\hline \hline \noalign{\smallskip}
Date &  BJD  &  Phase   &   Cycle     &       \multicolumn{1}{c}{$RV_{\mathrm c}$}     & \multicolumn{1}{c}{$ RV_{\mathrm m}$}        &   FWHM   & $D$  & \multicolumn{1}{c}{$A$}  &  S/N \\ 
\hline
2013.01.15 & 56307.778  &       0.00    &       52      &       -9.76   $_\mathrm{\pm   0.53    }$ &       -12.06$_\mathrm{\pm0.22}$  & 0.171      $_\mathrm{\pm   0.002   }$&     0.137   $_\mathrm{\pm   0.001   }$&     -15.9   $_\mathrm{\pm   2.2     }$ & 500   \\
2012.04.07 & 56024.531  &       0.03    &       44      &       -12.56  $_\mathrm{\pm   0.42    }$&     -13.01$_\mathrm{\pm     0.18    }$ & 0.155 $_\mathrm{\pm   0.001   }$&     0.154   $_\mathrm{\pm   0.001   }$&     -4.7    $_\mathrm{\pm   1.7     }$& 582     \\ 
2012.04.08 & 56025.514  &       0.06    &       44      &       -12.28  $_\mathrm{\pm   0.42    }$&      -12.97$_\mathrm{\pm     0.16    }$ & 0.150      $_\mathrm{\pm   0.001   }$&     0.168   $_\mathrm{\pm   0.001   }$&     -6.2    $_\mathrm{\pm   1.6     }$& 525     \\ 
2012.06.27 & 56105.505  &       0.31    &       46      &       0.43    $_\mathrm{\pm   0.32    }$&     -0.45$_\mathrm{\pm0.07  }$ & 0.112 $_\mathrm{\pm   0.001   }$&     0.288   $_\mathrm{\pm   0.001   }$&     -5.2    $_\mathrm{\pm   1.0     }$& 464     \\ 
2012.06.30 & 56109.468  &       0.42    &       46      &       7.79    $_\mathrm{\pm   0.54    }$&     6.37$_\mathrm{\pm       0.07    }$ & 0.121 $_\mathrm{\pm   0.001   }$&     0.292   $_\mathrm{\pm   0.001   }$&     -13.3   $_\mathrm{\pm   1.0     }$& 482     \\ 
2012.07.02 & 56111.486  &       0.47    &       46      &       11.47   $_\mathrm{\pm   0.55    }$&     9.91$_\mathrm{\pm       0.08    }$ &0.130  $_\mathrm{\pm   0.001   }$&     0.290   $_\mathrm{\pm   0.001   }$&     -15.9   $_\mathrm{\pm   1.1     }$& 428     \\ 
2013.01.01 & 56293.745  &       0.60    &       51      &       18.24   $_\mathrm{\pm   0.35    }$&     20.32$_\mathrm{\pm      0.10    }$ &0.157  $_\mathrm{\pm   0.001   }$&     0.253   $_\mathrm{\pm   0.001   }$&     15.5    $_\mathrm{\pm   1.1     }$& 340     \\ 
2012.05.01 & 56048.643  &       0.71    &       44      &       18.54   $_\mathrm{\pm   0.57    }$&     19.29$_\mathrm{\pm      0.13    }$ &0.183  $_\mathrm{\pm   0.001   }$&     0.227   $_\mathrm{\pm   0.001   }$&     3.7     $_\mathrm{\pm   1.0     }$& 395     \\ 
2012.05.04 & 56051.629  &       0.79    &       44      &       19.85   $_\mathrm{\pm   0.62    }$&     20.72$_\mathrm{\pm      0.15    }$ &0.201  $_\mathrm{\pm   0.001   }$&     0.204   $_\mathrm{\pm   0.001   }$&     4.2     $_\mathrm{\pm   1.2     }$& 324     \\ 
2012.04.02 & 56019.688  &       0.89    &       43      &       15.66   $_\mathrm{\pm   0.60    }$&     14.04$_\mathrm{\pm      0.21    }$ &0.202  $_\mathrm{\pm   0.002   }$&     0.153   $_\mathrm{\pm   0.001   }$&     -6.8    $_\mathrm{\pm   1.5     }$& 450     \\ 
2008.01.25 & 54491.266  &       0.90    &       1       &       14.56   $_\mathrm{\pm   0.57    }$&     12.21$_\mathrm{\pm      0.22    }$ &0.183  $_\mathrm{\pm   0.002   }$&     0.136   $_\mathrm{\pm   0.001   }$&     -14.0   $_\mathrm{\pm   2.0     }$& 453     \\ 
2008.01.26 & 54492.180  &       0.92    &       1       &       7.64    $_\mathrm{\pm   0.43    }$&     5.15$_\mathrm{\pm       0.20    }$   & 0.165       $_\mathrm{\pm   0.002   }$&     0.143   $_\mathrm{\pm   0.001   }$&     -17.1   $_\mathrm{\pm   2.2     }$& 424     \\ 
\hline \hline \noalign{\smallskip}
yyyy.mm.dd & BJD-2400000 &     &         &      \kms    &  \kms   &   \AA\   &  & \% \\ 
\hline
   
\end{tabular}
\end{center}
\end{table*}

\begin{figure}[htbp]
\begin{center}
\resizebox{0.6\hsize}{!}{\includegraphics[clip=true]{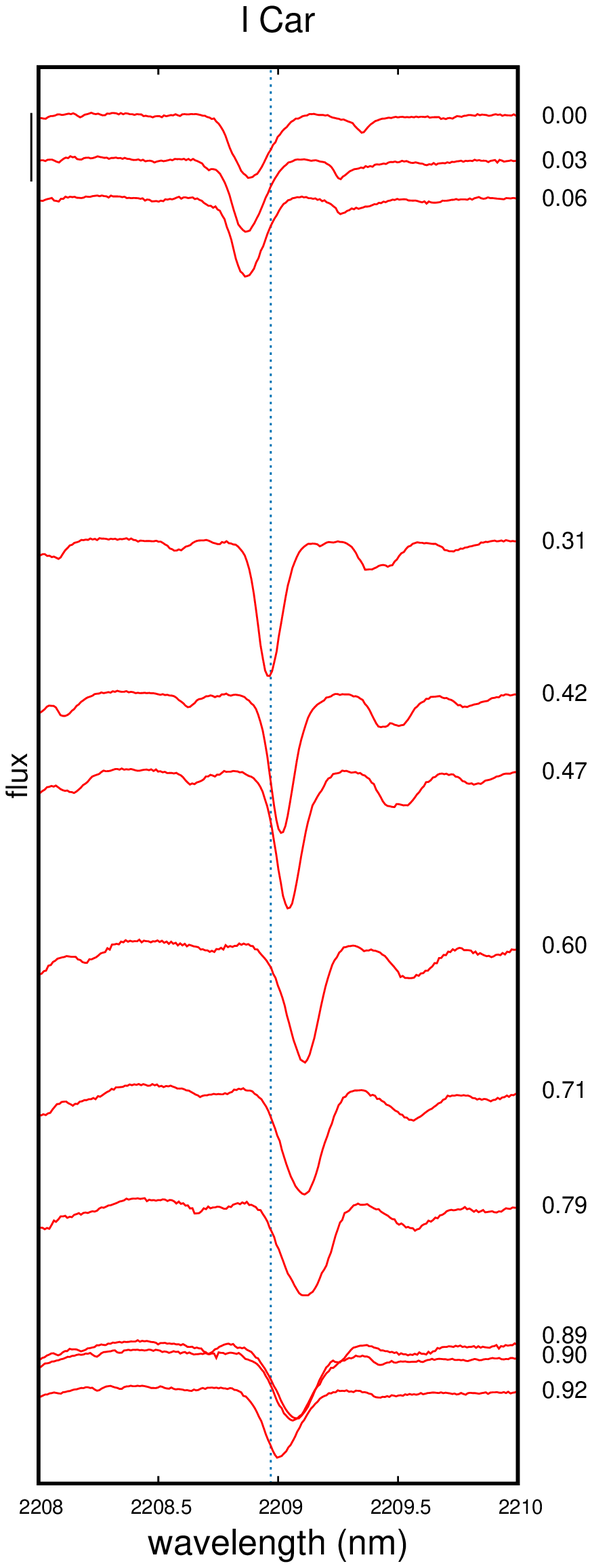}}
\end{center}
\vspace{-1cm}
\caption{Profile variations in the \ion{NaI} 2208.969~nm line obtained from the CRIRES spectra of \lcar\, (the reference wavelength is indicated by the blue vertical dotted line). The black vertical line at the top left corresponds to a differential flux of 0.15. The pulsation phases are indicated to the right of each profile. } \label{Fig.profile}
\end{figure}

\vspace{0.45cm}
\section{CRIRES observations of \lcar}\label{s_data}

The cryogenic high-resolution infrared echelle spectrograph CRIRES was located at the 
Nasmyth focus A of the UT1 of the ESO Very Large Telescope \citep{kaeufl04}. 
It provided a resolving power of up to 100,000 in the spectral range from 1 to 5.3~$\mu$m 
when used with a 0.2\arcsec\, slit.  The spectra were imaged on a detector mosaic composed 
of four Aladdin~III detectors (4096 x 512 pixel) with a gap of $\simeq$\,250 pixels between the chips. 
For all our observations, the spectral range is the same (in nm): [2178--2188], [2191--2201], [2204--2214], [2216--2226].

Our target was the long-period ($P$=35.551~d) Galactic Cepheid \lcar\,
(HD~84810$\equiv$HR~3884, variable from  $V$=3.35 to $V$=4.06, G5Iab/b).
 A detailed investigation of the period rate change over more than a century of observations is given by \citet{breitfelder16} (+27.283$\pm$0.984~s\,yr$^{-1}$) and \citet{neilson16} (+20.23$\pm$1.39~s\,yr$^{-1}$), who also concluded that \lcar\, does not show evidence  of the same enhanced mass-loss rates measured for $\delta$~Cep. However, a compact circumstellar envelope was detected by means of several instruments operating in the infrared 
\citep{kervella09}.

Most of the CRIRES spectra data 
were secured between April 2 and July 2, 2012.
Two spectra  were taken in January 2013, and we  added two taken in April 2008 \citep{nardetto11b}.
Table~\ref{Tab_res} reports the  log of observations. 
We used the EsoRex ESO Recipe Execution pipeline to reduce the data. However, the standard wavelength calibration of the spectra based on the thorium-argon lamp was unsuccessful (except for detector~1); therefore, we used the Molecfit ESO sky modeling tool in order to calibrate the data \citep{smette15, kausch15}\footnote{Thanks to 
the new and larger  CRIRES data set, we realized that two measurements (out of four) presented in  \citet{nardetto11b} (spectra at phase 0.27 and 0.78 in their Fig.~3; see also Sect. 4) were critically affected by an artifact in the wavelength calibration that could not be corrected. This led to an incorrect estimate of the infrared radial velocity amplitude and an incorrect line identification, while the general conclusions of the paper remain unchanged. These two  measurements are not used in the present study.}. For this, we compared each observed spectrum with its corresponding atmospheric model generated with Molecfit (including telluric lines) and we derived a mean offset of 10.7 $\pm$ 0.3 $\kms$. This value was used to calibrate the wavelength of all spectra.

By using the Atomic Spectra Bibliographic Databases of the National Institute of Standards and Technology (NIST) \citep{kramida16}, we were able to identify only one spectral line that was not blended by telluric lines at all phases, i.e., \ion{NaI} 2208.969~nm (in the wavelength range of detector 3). This line is the second component of a doublet;  the first component (\ion{NaI} 2206.242~nm), albeit within our spectral range of observation, is unfortunately blended by a telluric line, which prevents its analysis (Fig.~\ref{Fig.spectra}).

For the first time we explore the spectral region
beyond 2.0~$\mu$m;  the previous studies \citep{sasselov89,sasselov90} were performed in the 1.08--1.6~$\mu$m domain.
The line-profile changes are shown in Fig~\ref{Fig.profile}. 

From these spectral line profiles, we extract several physical quantities using the methods 
described in \citet{nardetto06a}. We fit a four-parameter analytic spectral line profile to 
the data \citep[Eqs.~2 and 3 in ][]{nardetto06a}: 
the depth of the line, $D$ (quantity without dimension);  
 the wavelength associated with the minimum of the line, $\lambda_{\mathrm m}$ (in \AA);  the full width at half maximum (FWHM) in the line (in \AA);
and  the line asymmetry, $A$ (in percentage of the FWHM).
All these parameters are listed in Table~\ref{Tab_res} together with the centroid velocity,
$RV_{\mathrm c}$, i.e., the first moment of each spectral line profile (Eq. 2 in \citet{nardetto06a}); 
$\lambda_{\mathrm m}$ values have been converted into velocities ($RV_{\mathrm m}$, not used 
in this study,  but listed for sake of completeness).

\begin{figure*}[htbp]
\begin{center}
\resizebox{0.85\hsize}{!}{\includegraphics[clip=true]{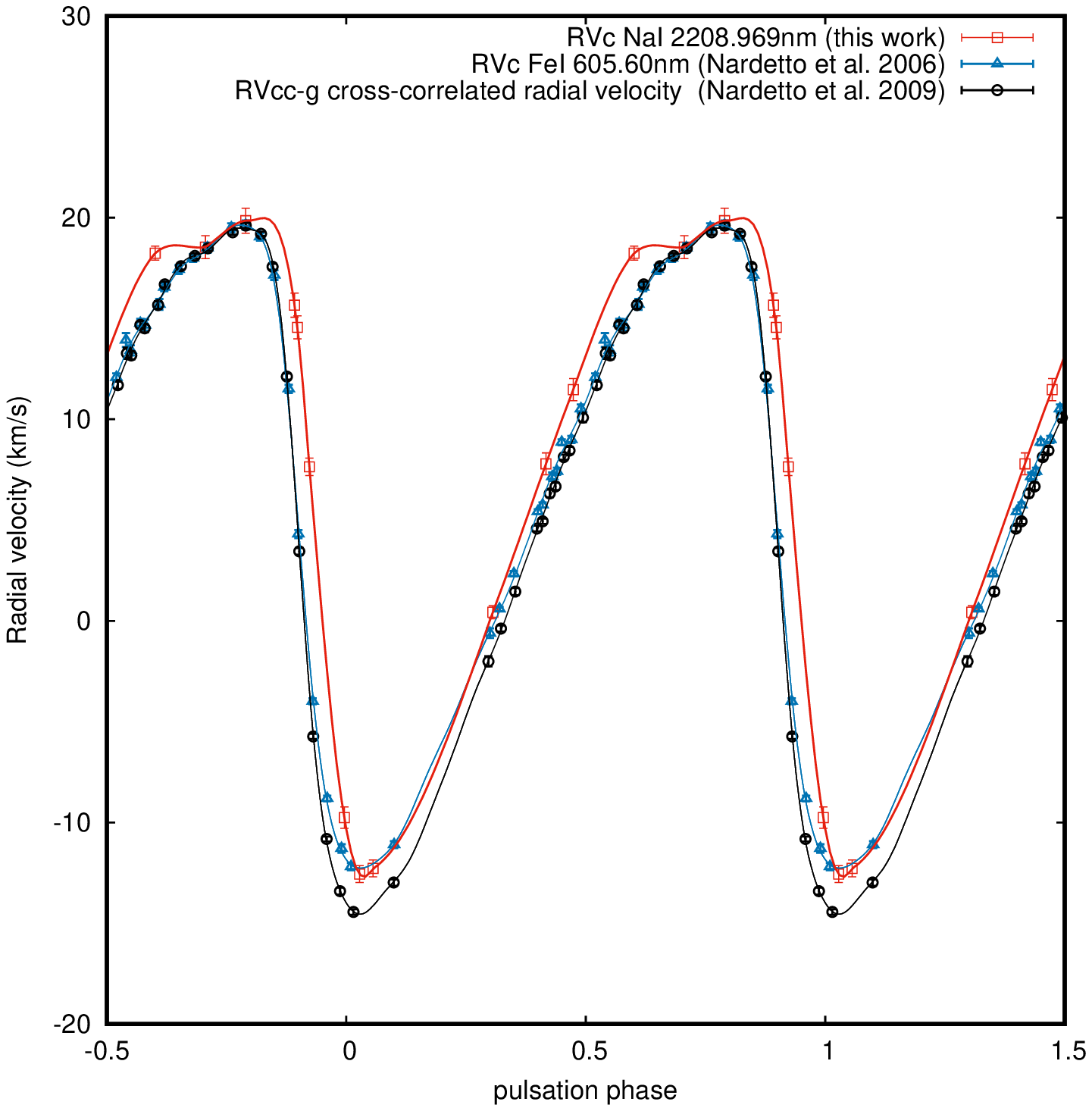}}
\end{center}
\caption{CRIRES infrared RV data (\ion{NaI} 2208.969~nm line, open red squares), HARPS optical RV data (line 605.6~nm, blue open triangles and fitting blue curve), and $RV_\mathrm{cc-g}$ data  (black open circles and fitting black line). 
} \label{Fig.RV}
\end{figure*}

\section{Comparison between infrared and visible observations}\label{s_analysis}

The similarity between the optical and infrared RV curves of Cepheids is not obvious. After the papers in the 1990s mentioned in Sect.~\ref{s_Introduction},
our HARPS and CRIRES data  of \lcar\, provide  one of the few examples currently available with modern instrumentation.

Taking into account the secular period variations \citep{neilson16}, we computed a period $P$=35.5496~d  and a time of maximum brightness $T_{\rm max}$=JD~2456023.533 at the epoch of the CRIRES observations.
The CRIRES $RV_{\mathrm c}$ data (red open squares in Fig.~\ref{Fig.RV}) folded with these elements  
show a time of maximum (receding) RV at BJD~2456053.05$\pm$1.0. The uncertainty is mostly due to fixing the real maximum RV phase and consequently the mean RV value. On the other hand, the minimum (approaching) RV is better fixed at BJD~2456025.0$\pm$0.5. The minimum RV corresponds to the maximum brightness $T_{\rm max}$, within a narrow phase lag between the two curves \citep{simon84}, so the agreement with the $T_{\rm max}$ reported above is very satisfactory.
We then apply a spline interpolation on these measurements (red solid line). 

For comparison purposes, we discuss the HARPS data presented in \citet{nardetto09}. We used the  $RV_\mathrm{cc-g}$ values, obtained
from the  Gaussian fit of the cross-correlation function.  We computed a period $P$=35.5476~d  and a $T_{\rm max}$=JD~2453001.573
at the epoch of these observations (January and February 2004). The $T_{\rm max}$ value is in excellent agreement with the observed minimum RV, i.e., BJD~2453002.46. We can  compare straightforwardly the  HARPS (visible) curve with the CRIRES (infrared) curve since the two ephemerides are based on the
periods and time of maxima  at the time of the respective observations. We also repeated the analysis by computing the phases assuming the same $T_0$ and an average period of 35.5486~d by obtaining very small phase shifts.

To make the comparison, we plot the $RV_{\mathrm c}$ curve associated with the line 605.6~nm presented in \citet{nardetto06a} (blue triangles in  Fig.~\ref{Fig.RV}). We also plot the  RV$_\mathrm{cc-g}$ values and curve used above (black empty circles and black solid line in Fig.~\ref{Fig.RV}). We recall  (see \citealt{nardetto09}) that the small differences between the RV curve of a line (like $RV_{\mathrm c}$ for   605.6~nm) and that  of the  $RV_\mathrm{cc-g}$ calculated from the whole spectrum is partly due to the $f_{\rm grad}$ component of the  projection factor $p$, and to the  way the RV values are derived (centroid versus Gaussian fit).

The bump seen at phases 0.47, 0.60, and 0.71 in the CRIRES RV might be due to a compression wave traveling in the upper part of the atmosphere 
\citep [probed by the extended forming regions of infrared lines; see Fig.~21 in ][]{sasselov90} or to  small variations in the shapes of the RV curves
 related to cycle 44 (phases 0.71 and 0.79), cycle 46 (phase 0.47), and cycle 51 (phase 0.60). Such cycle-to-cycle variations are reported in the RV curves of Cepheids \citep{anderson16a}.
We also note that the mean value of the CRIRES curve (red line in  Fig.~\ref{Fig.RV}) differs by 
$0.53\pm0.30$\,\kms\, from that of the HARPS  $RV_\mathrm{c}$ curve (blue line in Fig.~\ref{Fig.RV}). The uncertainty is due to wavelength calibration. Such an offset could be due to the different impact of granulation on the  line-forming regions in the optical and infrared. The deeper the line-forming region, the more the mean radial velocity is blueshifted \citep{vasilyev17}, which could indicate that the infrared line  forms in the upper part of the atmosphere. In order to verify this assumption, we compute  the layers (at the maximum radius) corresponding to an optical depth of $\frac{2}{3}$ at the center of the lines \ion{NaI} 2208.969~nm and \ion{FeI} 605.6~nm from the hydrodynamical model of \lcar\, presented in \citet{nardetto07}. We found 100\% and 20\%, respectively (0\% corresponds to the photospheric layer, while 100\% is the layer at the top of the atmosphere).  The core of the \ion{NaI} 2208.969~nm line  thus forms at the top of the atmosphere, conversely to the iron line in the visible.

\begin{figure}[htbp]
\begin{center}
\resizebox{0.9\hsize}{!}{\includegraphics[clip=true]{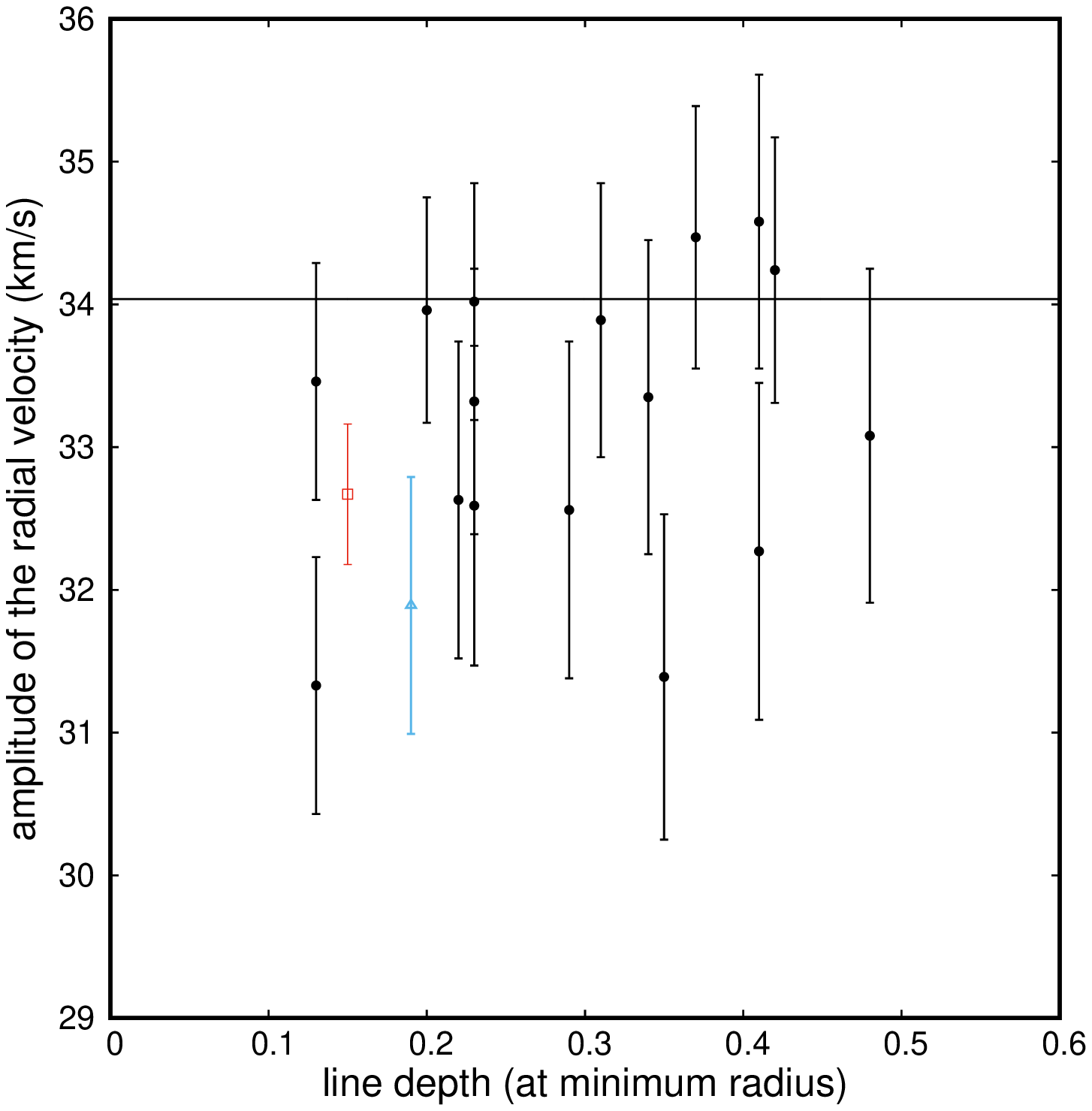}}
\end{center}
\caption{Amplitude of the $RV_{\mathrm c}$  curves  plotted as a function of the line depth for 17  different visible lines 
\citep{nardetto07} (black dots), including the 605.6~nm spectral line  (blue triangle), and the infrared line (red square). The horizontal black solid line is the amplitude of velocity associated with $RV_\mathrm{cc-g}$.} \label{Fig.grad}
\end{figure}

We note that the infrared $RV_{\mathrm c}$ curve has almost the same amplitude as its visible counterpart. This is seen in Fig.~\ref{Fig.grad} where we plot the amplitude of the $RV_{\mathrm c}$ curve associated with  the 17 lines (black dots), including 605.6~nm (blue triangles) of different depths.
The infrared line (red squares in Fig.~\ref{Fig.grad}) is consistent with the visible $RV_{\mathrm c}$ amplitudes. 
For comparison, the amplitude of the $RV_\mathrm{cc-g}$ (black solid line and circles in Fig.~\ref{Fig.RV}) 
is slightly larger, as already discussed in \citet{nardetto17}. 

\begin{figure}[htbp]
\begin{center}
\resizebox{0.9\hsize}{!}{\includegraphics[clip=true]{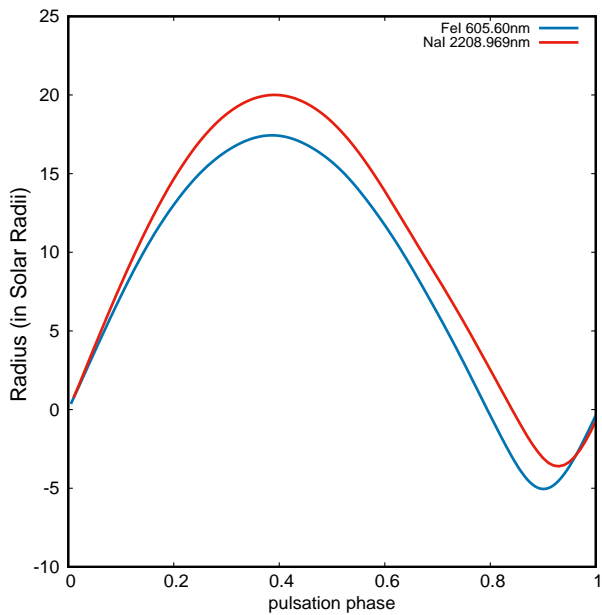}}
\end{center}
\caption{Interpolated radial velocity curves of Fig.~\ref{Fig.RV}  corrected from their respective velocity offsets and integrated with time in order to plot the radius variation associated with visible and infrared lines as a function of pulsation phase.} \label{Fig.R}
\end{figure}

\begin{figure}[htbp]
\begin{center}
\resizebox{0.9\hsize}{!}{\includegraphics[clip=true]{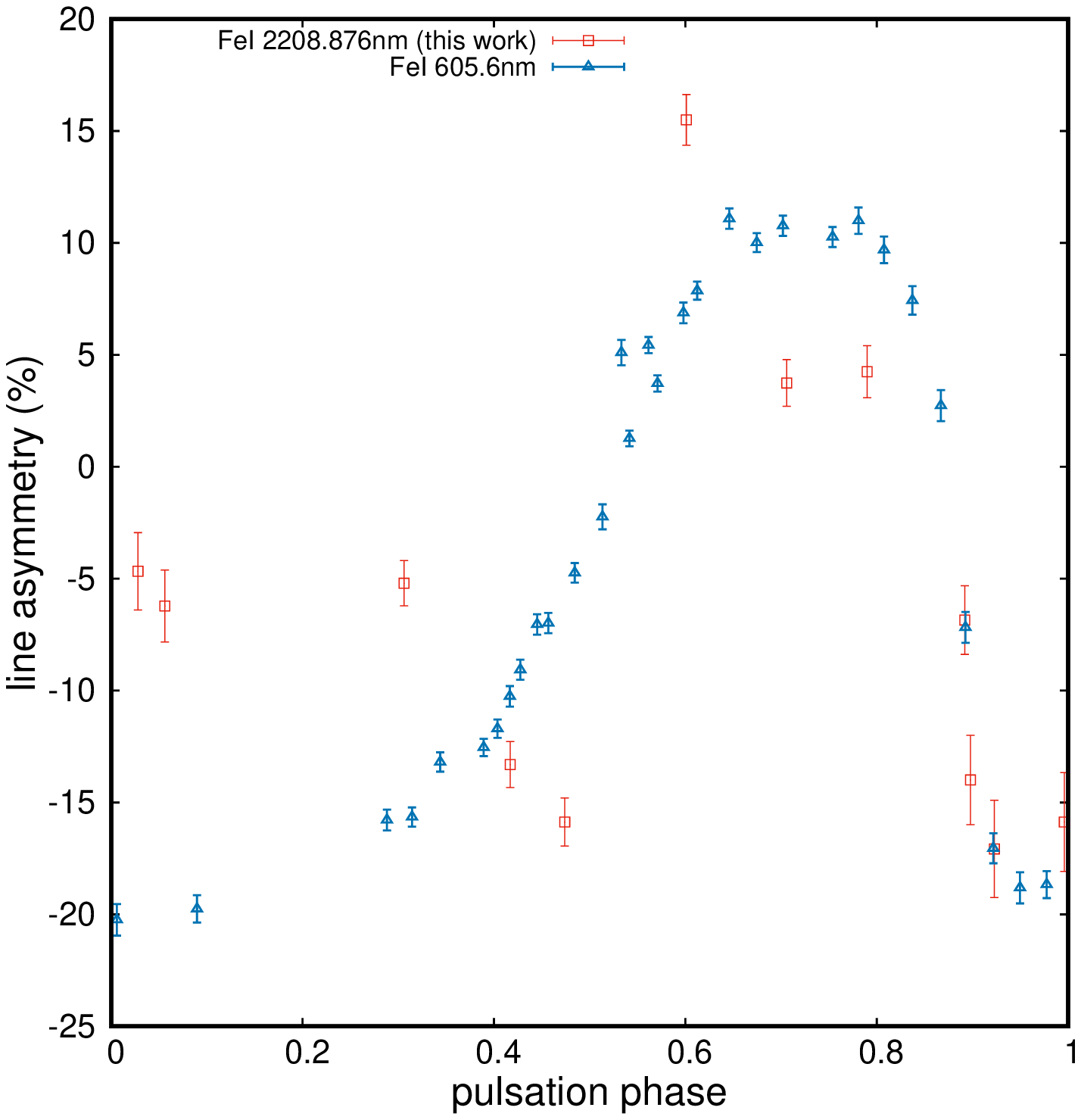}}
\end{center}
\caption{Infrared (red open squares) and visible (blue open triangles) line asymmetry curves, which are quite different. 
} \label{Fig.A}
\end{figure}

Another striking feature of the infrared radial velocity curve is the delay shift with respect to its visible counterpart that can be clearly seen in the descending branch. As proven by the use of different $P$ and $T_0$ values,  the shift
cannot be explained by the secular changes. 
The shift observed in the Cepheid \lcar\, is comparable to that observed in RR Lyrae stars by \citet{mathias95} between metallic and hydrogen lines in the visible (their Fig. 5), and related to the  Van Hoof effect \citep{vanhoof53}, i.e., 
 a delay in the velocities associated with the high and low line-forming regions. This effect has also been found in classical and type II Cepheids \citep{vinko98}. This can be seen in Fig.~\ref{Fig.R} where we plot the radius variation associated with  visible and infrared lines, respectively. This figure shows that a compression wave moves from the bottom to the top of the atmosphere during maximum outward velocity (phases around 0.9 to 1.0), which is consistent with Fig.~10 of \citet{sasselov94c} based on a time-dependent model of $\eta$~Aql.  This result seems to indicate, again, that the infrared line-forming region is probably located in the upper part of the atmosphere compared to the optical one, which might also explain  why the visible and infrared line asymmetry ($A$) curves are quite different in shape (Fig.~\ref{Fig.A}).  These differences in shape are probably due to different states of temperature and pressure between the bottom and top of the atmosphere, but also to different velocity fields (compression wave traveling). 




\section{Discussion and conclusion}\label{s_CL}

Our analysis of the behavior of the pulsation of \lcar\, in the infrared 
shows that the $RV_{\mathrm c}$ curve of the \ion{NaI} 2208.969~nm has about the same amplitude as the same curve obtained from optical lines, 
while a delay shift is noticeable in the descending branch (at minimum radius of the star). This observational result at 2.0~$\mu$m is different from that observed  
in the RV curves at 1.1~$\mu$m\, of the Cepheids X~Sgr and $\eta$~Aql, where amplitudes 20--35\% larger than
the optical values were measured \citep{sasselov89, sasselov90}. The same authors report on a doubling effect in the 1.08~$\mu$m line of short-period Cepheids X~Sgr, $\delta$~Cep, $\eta$~Aql, and possibly T~Mon, but we did not observe it in the 2.2~$\mu$m \ion{NaI}\, line of the long-period Cepheid \lcar\, (Fig.~\ref{Fig.profile}). Low [N/C] and [Na/Fe] ratios are reported in the atmosphere of \lcar\, \citep{luck11, genovali15}. Such peculiarities can play a role in the occurrence (or not) of atmospheric shockwaves. We recall that we did not find any line doubling in the visible range for $\delta$~Cep \citep{nardetto17} or for any Cepheid in our HARPS sample \citep{nardetto06a}. X~Sgr is the  only exception and the clear line-doubling was attributed to a double shockwave in the atmosphere \citep{mathias06}.

The observed equality between the two amplitudes is physically interesting. In principle, $RV_\mathrm{c}$ is independent of the width of the spectral line (rotation and micro-turbulence), while it depends on the pulsation velocity (associated with the line-forming region) and the limb-darkening. We refer to the limb darkening within the line and not to the limb darkening in the continuum, which is often used as an approximation  of the projection factor decomposition in the visible \citep{nardetto07}. Thus, equivalent visible and infrared $RV_\mathrm{c}$ amplitudes is consistent with similar pulsation velocities and limb darkenings (within the lines). Or conversely, it could indicate that the infrared line region is located in the upper part of the atmosphere as suggested by the hydrodynamical model,\footnote{As the hydrodynamical model is lacking a description of the convection, the detailed study of an infrared line in the case of {\mbox{l}~Car} located on the red side of the instability strip, is not recommended.} the delay shift (or Van Hoof effect), and the bump around phase 0.7 (dynamical effect or cycle-to-cycle variation). Thus, the pulsation velocity amplitude corresponding to the infrared line-forming region could be larger (compared to visible), which could compensate a weaker limb darkening within the infrared line.
In particular, the completely different behavior of the asymmetry of the  \ion{NaI} 2208.969~nm line 
also seems  to indicate that 
the infrared lines are sensitive to different physical processes with respect to optical lines and/or are formed at a different atmospheric height, which can also modify its RV velocity \citep{albrow96}.  Eventually, the impact of the granulation on the infrared line-forming region can contribute to the observed velocity shift between optical and infrared lines.

In any case, these results clearly indicate that the differences between infrared and optical high-resolution spectroscopy are a key to better understanding the dynamics  of the atmospheres of Cepheids. In particular, it could help in disclosing the true nature of the so-called $k$-term, i.e., the apparent systematic blueshifted motion at the level of a few \kms\ shown by all Cepheids \citep[e.g.,][]{nardetto09,vasilyev17}.  Simultaneous spectra acquired with new-generation, broad-range echelle spectrographs, like GIANO \citep{origlia14} and its updated version GIARPS \citep{claudi16} at the Telescopio Nazionale Galileo, or CRIRES+ at ESO \citep{follert14}, will be particularly helpful in this context.


\newpage
\begin{acknowledgements}


Based on observations collected at the European Organisation for Astronomical Research in the Southern Hemisphere
under ESO programme 089.D-0726(A). This research has made use of the SIMBAD and VIZIER\footnote{Available at http://cdsweb.u- strasbg.fr/} databases at CDS, Strasbourg (France), and of electronic bibliography maintained by the NASA/ADS system.   NN and EP acknowledge the {\it Observatoire de la C\^ote d'Azur} for the one-month grant that  allowed EP to work at OCA in June and July 2017. EP acknowledges financial support from PRIN INAF-2014. WG gratefully acknowledges financial support for this work from the BASAL Centro de Astrofisica y Tecnologias Afines (CATA) PFB-06/2007, and from the Millenium Institute of Astrophysics (MAS) of the Iniciativa Cientifica Milenio del Ministerio de Economia, Fomento y Turismo de Chile, project IC120009. Support from the Polish National Science Center grant MAESTRO 2012/06/A/ST9/00269 is also acknowledged. EP and MR acknowledge financial support from PRIN INAF-2014. NN, PK, AG, and WG acknowledge the support of the French--Chilean exchange program ECOS- Sud/CONICYT (C13U01). BP acknowledges financial support from the Polish National Science Center grant SONATA 2014/15/D/ST9/02248. The authors acknowledge the support of the French Agence Nationale de la Recherche (ANR), under grant ANR-15-CE31-0012- 01 (project UnlockCepheids) and the financial support from ``Programme National de Physique Stellaire'' (PNPS) of CNRS/INSU, France.

\end{acknowledgements}

\bibliographystyle{aa}  
\bibliography{bibtex_nn} 

\begin{thebibliography}{55}
\expandafter\ifx\csname natexlab\endcsname\relax\def\natexlab#1{#1}\fi

\bibitem[{{Albrow} \& {Cottrell}(1996)}]{albrow96}
{Albrow}, M.~D. \& {Cottrell}, P.~L. 1996, \mnras, 278, 337

\bibitem[{{Anderson}(2014)}]{anderson14}
{Anderson}, R.~I. 2014, \aap, 566, L10

\bibitem[{{Anderson}(2016)}]{anderson16a}
{Anderson}, R.~I. 2016, \mnras

\bibitem[{{Anderson} {et~al.}(2016){Anderson}, {M{\'e}rand}, {Kervella},
  {Breitfelder}, {LeBouquin}, {Eyer}, {Gallenne}, {Palaversa}, {Semaan},
  {Saesen}, \& {Mowlavi}}]{anderson16}
{Anderson}, R.~I., {M{\'e}rand}, A., {Kervella}, P., {et~al.} 2016, \mnras,
  455, 4231

\bibitem[{{Baade}(1926)}]{baade26}
{Baade}, W. 1926, Astronomische Nachrichten, 228, 359

\bibitem[{{Bell} \& {Rodgers}(1964)}]{bell64}
{Bell}, R.~A. \& {Rodgers}, A.~W. 1964, \mnras, 128, 365

\bibitem[{{Breitfelder} {et~al.}(2016){Breitfelder}, {M{\'e}rand}, {Kervella},
  {Gallenne}, {Szabados}, {Anderson}, \& {Le Bouquin}}]{breitfelder16}
{Breitfelder}, J., {M{\'e}rand}, A., {Kervella}, P., {et~al.} 2016, \aap, 587,
  A117

\bibitem[{{Claudi} {et~al.}(2016){Claudi}, {Benatti}, {Carleo}, {Ghedina},
  {Molinari}, {Oliva}, {Tozzi}, {Baruffolo}, {Cecconi}, {Cosentino},
  {Fantinel}, {Fini}, {Ghinassi}, {Gonzalez}, {Gratton}, {Guerra},
  {Harutyunyan}, {Hernandez}, {Iuzzolino}, {Lodi}, {Malavolta}, {Maldonado},
  {Micela}, {Sanna}, {Sanjuan}, {Scuderi}, {Sozzetti}, {P{\'e}rez Ventura},
  {Diaz Marcos}, {Galli}, {Gonzalez}, {Riverol}, \& {Riverol}}]{claudi16}
{Claudi}, R., {Benatti}, S., {Carleo}, I., {et~al.} 2016, in \procspie, Vol.
  9908, Ground-based and Airborne Instrumentation for Astronomy VI, 99081A

\bibitem[{{Davis} {et~al.}(2009){Davis}, {Jacob}, {Robertson}, {Ireland},
  {North}, {Tango}, \& {Tuthill}}]{davis09}
{Davis}, J., {Jacob}, A.~P., {Robertson}, J.~G., {et~al.} 2009, \mnras, 394,
  1620

\bibitem[{{Follert} {et~al.}(2014){Follert}, {Dorn}, {Oliva}, {Lizon},
  {Hatzes}, {Piskunov}, {Reiners}, {Seemann}, {Stempels}, {Heiter}, {Marquart},
  {Lockhart}, {Anglada-Escude}, {L{\"o}winger}, {Baade}, {Grunhut}, {Bristow},
  {Klein}, {Jung}, {Ives}, {Kerber}, {Pozna}, {Paufique}, {Kaeufl}, {Origlia},
  {Valenti}, {Gojak}, {Hilker}, {Pasquini}, {Smette}, \& {Smoker}}]{follert14}
{Follert}, R., {Dorn}, R.~J., {Oliva}, E., {et~al.} 2014, in \procspie, Vol.
  9147, Ground-based and Airborne Instrumentation for Astronomy V, 914719

\bibitem[{{Fouqu{\'e}} {et~al.}(2007){Fouqu{\'e}}, {Arriagada}, {Storm},
  {Barnes}, {Nardetto}, {M{\'e}rand}, {Kervella}, {Gieren}, {Bersier},
  {Benedict}, \& {McArthur}}]{fouque07}
{Fouqu{\'e}}, P., {Arriagada}, P., {Storm}, J., {et~al.} 2007, \aap, 476, 73

\bibitem[{{Fouque} \& {Gieren}(1997)}]{fouque97}
{Fouque}, P. \& {Gieren}, W.~P. 1997, \aap, 320, 799

\bibitem[{{Genovali} {et~al.}(2015){Genovali}, {Lemasle}, {da Silva}, {Bono},
  {Fabrizio}, {Bergemann}, {Buonanno}, {Ferraro}, {Fran{\c c}ois}, {Iannicola},
  {Inno}, {Laney}, {Kudritzki}, {Matsunaga}, {Nonino}, {Primas}, {Romaniello},
  {Urbaneja}, \& {Th{\'e}venin}}]{genovali15}
{Genovali}, K., {Lemasle}, B., {da Silva}, R., {et~al.} 2015, \aap, 580, A17

\bibitem[{{Kaeufl} {et~al.}(2004){Kaeufl}, {Ballester}, {Biereichel},
  {Delabre}, {Donaldson}, {Dorn}, {Fedrigo}, {Finger}, {Fischer}, {Franza},
  {Gojak}, {Huster}, {Jung}, {Lizon}, {Mehrgan}, {Meyer}, {Moorwood}, {Pirard},
  {Paufique}, {Pozna}, {Siebenmorgen}, {Silber}, {Stegmeier}, \&
  {Wegerer}}]{kaeufl04}
{Kaeufl}, H.-U., {Ballester}, P., {Biereichel}, P., {et~al.} 2004, in Society
  of Photo-Optical Instrumentation Engineers (SPIE) Conference Series, Vol.
  5492, Ground-based Instrumentation for Astronomy, ed. A.~F.~M. {Moorwood} \&
  M.~{Iye}, 1218--1227

\bibitem[{{Karp}(1975)}]{karp75c}
{Karp}, A.~H. 1975, \apj, 201, 641

\bibitem[{{Kausch} {et~al.}(2015){Kausch}, {Noll}, {Smette}, {Kimeswenger},
  {Barden}, {Szyszka}, {Jones}, {Sana}, {Horst}, \& {Kerber}}]{kausch15}
{Kausch}, W., {Noll}, S., {Smette}, A., {et~al.} 2015, \aap, 576, A78

\bibitem[{{Kervella} {et~al.}(2004{\natexlab{a}}){Kervella}, {Fouqu{\'e}},
  {Storm}, {Gieren}, {Bersier}, {Mourard}, {Nardetto}, \& {du Coud{\'e}
  Foresto}}]{kervella04d}
{Kervella}, P., {Fouqu{\'e}}, P., {Storm}, J., {et~al.} 2004{\natexlab{a}},
  \apjl, 604, L113

\bibitem[{{Kervella} {et~al.}(2009){Kervella}, {M{\'e}rand}, \&
  {Gallenne}}]{kervella09}
{Kervella}, P., {M{\'e}rand}, A., \& {Gallenne}, A. 2009, \aap, 498, 425

\bibitem[{{Kervella} {et~al.}(2004{\natexlab{b}}){Kervella}, {Nardetto},
  {Bersier}, {Mourard}, \& {Coud{\'e} du Foresto}}]{kervella04a}
{Kervella}, P., {Nardetto}, N., {Bersier}, D., {Mourard}, D., \& {Coud{\'e} du
  Foresto}, V. 2004{\natexlab{b}}, \aap, 416, 941

\bibitem[{{Kramida} {et~al.}(2016){Kramida}, {Ralchenko}, \&
  {Reader}}]{kramida16}
{Kramida}, A., {Ralchenko}, Y., \& {Reader}, J. 2016, in APS Division of
  Atomic, Molecular and Optical Physics Meeting Abstracts

\bibitem[{{Kurucz}(1979)}]{kurucz79}
{Kurucz}, R.~L. 1979, \apjs, 40, 1

\bibitem[{{Lane} {et~al.}(2000){Lane}, {Kuchner}, {Boden}, {Creech-Eakman}, \&
  {Kulkarni}}]{lane00}
{Lane}, B.~F., {Kuchner}, M.~J., {Boden}, A.~F., {Creech-Eakman}, M., \&
  {Kulkarni}, S.~R. 2000, \nat, 407, 485

\bibitem[{{Lindemann}(1918)}]{lindermann18}
{Lindemann}, F.~A. 1918, \mnras, 78, 639

\bibitem[{{Luck} \& {Lambert}(2011)}]{luck11}
{Luck}, R.~E. \& {Lambert}, D.~L. 2011, \aj, 142, 136

\bibitem[{{Maillard} \& {Michel}(1982)}]{maillard82}
{Maillard}, J.~P. \& {Michel}, G. 1982, in Astrophysics and Space Science
  Library, Vol.~92, IAU Colloq. 67: Instrumentation for Astronomy with Large
  Optical Telescopes, ed. C.~M. {Humphries}, 213--222

\bibitem[{{Mathias} {et~al.}(1995){Mathias}, {Gillet}, {Fokin}, \&
  {Chadid}}]{mathias95}
{Mathias}, P., {Gillet}, D., {Fokin}, A.~B., \& {Chadid}, M. 1995, \aap, 298,
  843

\bibitem[{{Mathias} {et~al.}(2006){Mathias}, {Gillet}, {Fokin}, {Nardetto},
  {Kervella}, \& {Mourard}}]{mathias06}
{Mathias}, P., {Gillet}, D., {Fokin}, A.~B., {et~al.} 2006, \aap, 457, 575

\bibitem[{{Merand} {et~al.}(2015){Merand}, {Kervella}, {Breitfelder},
  {Gallenne}, {Coude du Foresto}, {ten Brummelaar}, {McAlister}, {Ridgway},
  {Sturmann}, {Sturmann}, \& {Turner}}]{merand15}
{Merand}, A., {Kervella}, P., {Breitfelder}, J., {et~al.} 2015, ArXiv e-prints

\bibitem[{{M{\'e}rand} {et~al.}(2005){M{\'e}rand}, {Kervella}, {Coud{\'e} du
  Foresto}, {Ridgway}, {Aufdenberg}, {ten Brummelaar}, {Berger}, {Sturmann},
  {Sturmann}, {Turner}, \& {McAlister}}]{merand05}
{M{\'e}rand}, A., {Kervella}, P., {Coud{\'e} du Foresto}, V., {et~al.} 2005,
  \aap, 438, L9

\bibitem[{{Nardetto} {et~al.}(2011){Nardetto}, {Fokin}, {Fouqu{\'e}}, {Storm},
  {Gieren}, {Pietrzynski}, {Mourard}, \& {Kervella}}]{nardetto11b}
{Nardetto}, N., {Fokin}, A., {Fouqu{\'e}}, P., {et~al.} 2011, \aap, 534, L16

\bibitem[{{Nardetto} {et~al.}(2004){Nardetto}, {Fokin}, {Mourard}, {Mathias},
  {Kervella}, \& {Bersier}}]{nardetto04}
{Nardetto}, N., {Fokin}, A., {Mourard}, D., {et~al.} 2004, \aap, 428, 131

\bibitem[{{Nardetto} {et~al.}(2009){Nardetto}, {Gieren}, {Kervella},
  {Fouqu{\'e}}, {Storm}, {Pietrzynski}, {Mourard}, \& {Queloz}}]{nardetto09}
{Nardetto}, N., {Gieren}, W., {Kervella}, P., {et~al.} 2009, \aap, 502, 951

\bibitem[{{Nardetto} {et~al.}(2016){Nardetto}, {M{\'e}rand}, {Mourard},
  {Storm}, {Gieren}, {Fouqu{\'e}}, {Gallenne}, {Graczyk}, {Kervella},
  {Neilson}, {Pietrzynski}, {Pilecki}, {Breitfelder}, {Berio}, {Challouf},
  {Clausse}, {Ligi}, {Mathias}, {Meilland}, {Perraut}, {Poretti}, {Rainer},
  {Spang}, {Stee}, {Tallon-Bosc}, \& {ten Brummelaar}}]{nardetto16a}
{Nardetto}, N., {M{\'e}rand}, A., {Mourard}, D., {et~al.} 2016, \aap, 593, A45

\bibitem[{{Nardetto} {et~al.}(2006){Nardetto}, {Mourard}, {Kervella},
  {Mathias}, {M{\'e}rand}, \& {Bersier}}]{nardetto06a}
{Nardetto}, N., {Mourard}, D., {Kervella}, P., {et~al.} 2006, \aap, 453, 309

\bibitem[{{Nardetto} {et~al.}(2007){Nardetto}, {Mourard}, {Mathias}, {Fokin},
  \& {Gillet}}]{nardetto07}
{Nardetto}, N., {Mourard}, D., {Mathias}, P., {Fokin}, A., \& {Gillet}, D.
  2007, \aap, 471, 661

\bibitem[{{Nardetto} {et~al.}(2017){Nardetto}, {Poretti}, {Rainer}, {Fokin},
  {Mathias}, {Anderson}, {Gallenne}, {Gieren}, {Graczyk}, {Kervella},
  {M{\'e}rand}, {Mourard}, {Neilson}, {Pietrzynski}, {Pilecki}, \&
  {Storm}}]{nardetto17}
{Nardetto}, N., {Poretti}, E., {Rainer}, M., {et~al.} 2017, \aap, 597, A73

\bibitem[{{Neilson} {et~al.}(2016){Neilson}, {Engle}, {Guinan}, {Bisol}, \&
  {Butterworth}}]{neilson16}
{Neilson}, H.~R., {Engle}, S.~G., {Guinan}, E.~F., {Bisol}, A.~C., \&
  {Butterworth}, N. 2016, \apj, 824, 1

\bibitem[{{Neilson} {et~al.}(2012){Neilson}, {Nardetto}, {Ngeow}, {Fouqu{\'e}},
  \& {Storm}}]{neilson12}
{Neilson}, H.~R., {Nardetto}, N., {Ngeow}, C.-C., {Fouqu{\'e}}, P., \& {Storm},
  J. 2012, \aap, 541, A134

\bibitem[{{Ngeow} {et~al.}(2012){Ngeow}, {Neilson}, {Nardetto}, \&
  {Marengo}}]{ngeow12}
{Ngeow}, C.-C., {Neilson}, H.~R., {Nardetto}, N., \& {Marengo}, M. 2012, \aap,
  543, A55

\bibitem[{{Origlia} {et~al.}(2014){Origlia}, {Oliva}, {Baffa}, {Falcini},
  {Giani}, {Massi}, {Montegriffo}, {Sanna}, {Scuderi}, {Sozzi}, {Tozzi},
  {Carleo}, {Gratton}, {Ghinassi}, \& {Lodi}}]{origlia14}
{Origlia}, L., {Oliva}, E., {Baffa}, C., {et~al.} 2014, in \procspie, Vol.
  9147, Ground-based and Airborne Instrumentation for Astronomy V, 91471E

\bibitem[{{Sanford}(1956)}]{sanford56}
{Sanford}, R.~F. 1956, \apj, 123, 201

\bibitem[{{Sasselov} {et~al.}(1989){Sasselov}, {Fieldus}, \&
  {Lester}}]{sasselov89}
{Sasselov}, D.~D., {Fieldus}, M.~S., \& {Lester}, J.~B. 1989, \apjl, 337, L29

\bibitem[{{Sasselov} \& {Lester}(1990)}]{sasselov90}
{Sasselov}, D.~D. \& {Lester}, J.~B. 1990, \apj, 362, 333

\bibitem[{{Sasselov} \& {Lester}(1994{\natexlab{a}})}]{sasselov94c}
{Sasselov}, D.~D. \& {Lester}, J.~B. 1994{\natexlab{a}}, \apj, 423, 795

\bibitem[{{Sasselov} \& {Lester}(1994{\natexlab{b}})}]{sasselov94a}
{Sasselov}, D.~D. \& {Lester}, J.~B. 1994{\natexlab{b}}, \apj, 423, 777

\bibitem[{{Sasselov} \& {Lester}(1994{\natexlab{c}})}]{sasselov94b}
{Sasselov}, D.~D. \& {Lester}, J.~B. 1994{\natexlab{c}}, \apj, 423, 785

\bibitem[{{Simon}(1984)}]{simon84}
{Simon}, N.~R. 1984, \apj, 284, 278

\bibitem[{{Smette} {et~al.}(2015){Smette}, {Sana}, {Noll}, {Horst}, {Kausch},
  {Kimeswenger}, {Barden}, {Szyszka}, {Jones}, {Gallenne}, {Vinther},
  {Ballester}, \& {Taylor}}]{smette15}
{Smette}, A., {Sana}, H., {Noll}, S., {et~al.} 2015, \aap, 576, A77

\bibitem[{{Storm} {et~al.}(2011{\natexlab{a}}){Storm}, {Gieren}, {Fouqu{\'e}},
  {Barnes}, {Pietrzy{\'n}ski}, {Nardetto}, {Weber}, {Granzer}, \&
  {Strassmeier}}]{storm11a}
{Storm}, J., {Gieren}, W., {Fouqu{\'e}}, P., {et~al.} 2011{\natexlab{a}}, \aap,
  534, A94

\bibitem[{{Storm} {et~al.}(2011{\natexlab{b}}){Storm}, {Gieren}, {Fouqu{\'e}},
  {Barnes}, {Soszy{\'n}ski}, {Pietrzy{\'n}ski}, {Nardetto}, \&
  {Queloz}}]{storm11b}
{Storm}, J., {Gieren}, W., {Fouqu{\'e}}, P., {et~al.} 2011{\natexlab{b}}, \aap,
  534, A95

\bibitem[{{van Hoof} \& {Struve}(1953)}]{vanhoof53}
{van Hoof}, A. \& {Struve}, O. 1953, \pasp, 65, 158

\bibitem[{{Vasilyev} {et~al.}(2017){Vasilyev}, {Ludwig}, {Freytag}, {Lemasle},
  \& {Marconi}}]{vasilyev17}
{Vasilyev}, V., {Ludwig}, H.-G., {Freytag}, B., {Lemasle}, B., \& {Marconi}, M.
  2017, ArXiv e-prints

\bibitem[{{Vinko} {et~al.}(1998){Vinko}, {Remage Evans}, {Kiss}, \&
  {Szabados}}]{vinko98}
{Vinko}, J., {Remage Evans}, N., {Kiss}, L.~L., \& {Szabados}, L. 1998, \mnras,
  296, 824

\bibitem[{{Wallerstein} {et~al.}(2015){Wallerstein}, {Albright}, \&
  {Ritchey}}]{wallerstein15}
{Wallerstein}, G., {Albright}, M.~B., \& {Ritchey}, A.~M. 2015, \pasp, 127, 503

\bibitem[{{Wesselink}(1946)}]{wesselink46}
{Wesselink}, A.~J. 1946, \bain, 10, 91

\end{thebibliography}

\end{document}